\documentclass[12pt]{article}

\def\thefootnote{\fnsymbol{footnote}}

\usepackage{latexsym, epsf}

\begin{document}

\newcommand{\gtrsim}{ \mathop{}_{\textstyle \sim}^{\textstyle >} }
\newcommand{\lesssim}{ \mathop{}_{\textstyle \sim}^{\textstyle <} }

\begin{titlepage}

\begin{center}

\hfill  ICRR-Report-374-96-25\\
\hfill  UT-758 \\
\hfill  hep-ph/9608405\\
\vspace{1.5cm}

{\large\textbf{Cosmological Axion Problem in Chaotic Inflationary Universe}}

\vspace{1cm}

{\large S.~Kasuya$^a$\footnote{kasuya@icrr.u-tokyo.ac.jp}, 
M.~Kawasaki$^a$\footnote{kawasaki@icrr.u-tokyo.ac.jp} 
and T.~Yanagida$^b$\footnote{yanagida@danjuro.phys.s.u-tokyo.ac.jp}} \\

\vspace{1cm}

$^a$\textit{Institute for Cosmic Ray Research, University of Tokyo,
  Tanashi, Tokyo 188, Japan} \\
$^b$\textit{Department of Physics, School of Science, University of
  Tokyo, Tokyo 113, Japan}
%%
%%
%%\date{\today}
 
\end{center}

\vspace{1cm}

\begin{abstract}
    We investigate two cosmological axion problems ( isocurvature
    fluctuations and domain-wall formation ) in chaotic inflationary
    universe. It is believed that these problems are solved if
    potential for the Peccei-Quinn scalar field is very flat. However,
    we find that too many domain walls are produced through parametric
    resonance decay of the Peccei-Quinn scalar field. Only the axion
    model with $N=1$($N$: QCD anomaly factor) is consistent with
    observations. We also point out that the flat potential is
    naturally obtained in a supersymmetric extension of the
    Peccei-Quinn model. If Peccei-Quinn breaking scale $F_a$ is about
    $10^{12}$~GeV, this model predicts anisotropies of cosmic
    microwave background radiation due to the axion isocurvature
    fluctuations which may be detectable in future observations.
\end{abstract}

\end{titlepage}

\newpage

\addtocounter{footnote}{-3}
\renewcommand{\thefootnote}{\arabic{footnote}}

\section{Introduction}

The axion~\cite{Peccei,Wilczek,Kim,DFS} is a Nambu-Goldstone boson
associated with breaking of the Peccei-Quinn symmetry which was
invented as a solution to the strong CP problem in QCD~\cite{tHoot}.
The Peccei-Quinn symmetry breaking scale $F_a$ is stringently
constrained by laboratory experiments, astrophysics and cosmology:
allowed range of $F_a$ lies between $10^{10}$~GeV and
$10^{12}$~GeV~\cite{Kolb-Turner}. The axion is also cosmologically
attractive since it can be a cold dark matter if $F_a$ takes the
higher values $\sim 10^{12}$~GeV.

Inflationary universe~\cite{Guth,Sato} was proposed to solve various
problems in the standard cosmology (e.g. flatness and horizon
problems). There have been constructed many inflation models since
then. The chaotic inflation model~\cite{Linde1} seems the most
attractive candidate among them that realizes naturally the
inflationary universe. However, when we consider the axion in the
chaotic inflationary universe, we are confronted with two serious
cosmological problems, both of which are associated with large quantum
fluctuations generated in the exponentially expanding phase (i.e.,
inflationary epoch) of the early universe. One is the domain wall
problem~\cite{Lyth1}. In the inflationary universe fluctuations of the
axion field $a(x)$ are given by $\delta a = H/(2\pi)$ where $H$ is the
Hubble constant. Since a phase $\theta_a$ of the Peccei-Quinn scalar
field is related to the axion field $a(x)$ by $\theta_a = a/F_a$, the
fluctuations of $\theta_a$ is given by
\begin{equation}
    \label{delta-theta}
    \delta \theta_a = \frac{H}{2\pi F_a}.
\end{equation}
In the chaotic inflation model, $H \simeq 10^{14}$~GeV~\cite{Salopek}
is required to explain the anisotropies of cosmic microwave background
radiation (CBR) observed by COBE~\cite{COBE}. Then, the fluctuations
of the phase $\theta_a$ become $O(1)$ for $F_a \lesssim 10^{12}$~GeV,
which means that the phase is quite random during the inflation.
Therefore, when the universe cools down to about 1~GeV and the axion
potential is formed, the axion sits at different positions of the
potential in different regions of the universe. Since the axion
potential has $N$ discrete minima ($N$: QCD anomaly factor), domain
walls are produced~\cite{Sikivie}. The domain wall with $N=1$ is a
disk-like object whose boundary is an axionic string and it collapses
quickly due to its surface tension. Thus the domain wall with $N=1$ is
cosmologically harmless. However, the domain wall with $N \ge 2$ is
disastrous because it forms a complicated network with an axionic
string and dominates the energy density of the universe
quickly~\cite{Ryden}.

The second problem is that the quantum fluctuations for the axion
field cause too large anisotropies of
CBR~\cite{Turner,Lyth2,LiLy,Linde2,KSY}.  Since the axion does not
have a potential during the inflation, the axion fluctuations $\delta
a$ do not contribute to the energy density of the universe. In this
sense, the axion fluctuations are isocurvature.  After the axion
acquires a mass $m_a$, its fluctuations become density fluctuations
given by $\delta \rho_a/\rho_a \sim \delta \theta_a/\theta_a$ which
causes the CBR temperature fluctuations $\delta T/T \sim \delta
\theta_a/\theta_a$.  From eq.(\ref{delta-theta}), we see that the
produced CBR anisotropies are $O(1)$ which contradicts the
observation.

It has been pointed out in ref.\cite{LiLy,Linde2} that these problems
are simultaneously solved if the potential for the Peccei-Quinn scalar
field is very flat. In this case it is natural that the Peccei-Quinn
scalar field $\Phi$ takes a very large value $\sim M_{\rm pl}$ during
the inflation. Here, $M_{\rm pl}$ is the Planck mass ($M_{\rm pl} =
1.2\times 10^{19}$~GeV). The axion field $a(x)$ is defined as
\begin{eqnarray}
    \Phi(x) & \equiv &  \phi(x)
    \exp(i a(x)/|\langle\phi\rangle|),\\
    & & \phi(x): {\rm real},~ -\infty < \phi < \infty, \nonumber \\
    & & a(x):{\rm real},~ -
    \pi/2 \le a(x)/|\langle\phi\rangle| \le \pi/2. \nonumber 
\end{eqnarray}
In the true vacuum $\phi$ takes $ |\langle\phi\rangle| = F_a$.  But
the $\phi$ takes most likely $|\langle \phi \rangle| \sim M_{\rm pl}$
at the inflationary epoch. From eq.(\ref{delta-theta}) it is clear
that the phase fluctuations $\delta \theta_a$ are suppressed as
$\delta\theta_a \simeq H/(2\pi M_{\rm pl})$. Therefore the
isocurvature fluctuations are suppressed by a factor $F_a/M_{\rm pl}$.
In ref.~\cite{KSY}, cosmological effects of the axionic isocurvature
density fluctuations have been investigated in details by calculating
spectra for matter fluctuations and $\delta T/T$, and it has been
found that the effects of the isocurvature fluctuations are consistent
with observations in a large parameter region.

As for the domain wall problem, it is naively expected that the same
mechanism that suppresses the isocurvature fluctuations does also
suppress the production of domain walls: i.e. the large effective
``$F_a$''$ \equiv |\langle \phi \rangle|_{\rm inflation} \sim M_{\rm
pl}$ during the inflation almost fixes the phase of $\Phi$ which
reduces the domain-wall production rate to negligible amount. In fact,
the $\Phi$ begins to oscillate after the inflation with its phase
being fixed. However, the homogeneous Peccei-Quinn field decays into
axions as well as $\phi$ particles through parametric resonances. The
emitted axions induce large fluctuations of the axion field (i.e. the
phase $\delta \theta_a$) and produce a large number of domain walls.

In this letter we study the above cosmological axion problems (
isocurvature-fluctuation and domain-wall problems ) in chaotic
inflationary universe.  We show that the domain walls are still
produced through the parametric resonance decay even if we take a very
flat potential for the Peccei-Quinn scalar. The domain wall problem is
only avoided for the model with $N=1$. Furthermore, we also show that
the flat potential for $\Phi$ required to solve the
isocurvature-fluctuation problem is naturally obtained in a
supersymmetric (SUSY) extension of the Peccei-Quinn model. Namely, the
presence of flat directions is a generic feature in SUSY theories and
the masses of the fields corresponding to the flat directions only
comes from soft SUSY breaking terms which are of order the weak scale.
Therefore the flat potential for $\Phi$ is naturally realized in the
framework of SUSY.

\section{Domain Wall Problem for a Quartic Potential}
%\label{sec:wall}

Let us first consider the case for a scalar potential given by
\begin{equation}
    V(\Phi) = \frac{g}{4} (|\Phi|^2 -F_a^2)^2
     = \frac{g}{4} (\phi^2 -F_a^2)^2,
    \label{potential}
\end{equation}
with $g \ll 1$.  In the inflationary epoch, the $\phi$ slowly evolves
as
\begin{equation}
    \phi \simeq \left(\frac{\lambda}{g}\right)^{1/2} \chi,
    \label{Phi_a-phi}
\end{equation}
where we have assumed that the potential for the inflaton $\chi$ is
$\lambda \chi^4/4$. Isocurvature fluctuations for cosmologically
relevant scales are generated when $\chi \simeq 4M_{\rm pl}$.
Therefore the CBR anisotropies from the axion
isocurvature fluctuations is given by~\footnote{%%
For a more accurate analysis including comparison with the most recent
observations, see ref.~\cite{KSY}}
\begin{equation}
    \frac{\delta T}{T} \sim \frac{\delta \rho_a}{\rho_a} 
    \sim \sqrt{\frac{2}{3\pi}}g^{1/2}
    \left(\frac{\chi}{M_{\rm pl}\theta_a}\right)
    \simeq 2 \frac{g^{1/2}}{\theta_a}.
    \label{isocrv}
\end{equation}
Using a COBE constraint $\delta T/T \lesssim 10^{-5}$,  we get $g
\lesssim 10^{-11}$ for $\theta_a \sim O(1)$.~\footnote{%%
Even if one assumes the inflaton potential $\sim m_{\chi}^2\chi^2/2$,
one gets essentially the same conclusion as in the text.}
Notice that the coupling constant $\lambda$ for the inflaton is already very
small as $\lambda \sim 10^{-13}$~\cite{Salopek}.

Next we consider domain wall production with the potential
(\ref{potential}). The Peccei-Quinn field $\Phi$ starts to oscillate
after the inflation.  We only consider the evolution of $\Phi$ in the
``radial'' direction $\phi(x)$ with its phase being fixed since the
quantum fluctuations of the axion phase $\delta \theta_a\equiv \delta
a/\langle \phi \rangle$ are negligiblly small when $|\langle \phi
\rangle| \sim M_{\rm pl}$ as discussed in the introduction. As easily
seen from eq.(\ref{potential}) the ``radial'' potential has two minima
($\phi =
\pm F_a $).~\footnote{%%
Note that in the conventional notation of the axion field $\Phi =
|\langle\Phi\rangle|e^{ia(x)/|\langle\Phi\rangle|}$ the two minima
correspond to $\theta_a^0$ and $\theta_a^0+\pi$ where $\theta_a^0$ is
the initial value of the axion phase.}
Since the initial potential energy density of $\phi$ is large, the
$\phi$ oscillates beyond the potential hill at $\phi = 0$. The $\phi$
loses its energy through cosmic expansion and particle production and
settles down to one of the potential minima eventually.

If the dynamics after inflation is purely classical, the final value
of $\phi$ ($F_a$ or $-F_a$) depends on the initial value of $\phi$.
However, it has been recently pointed out that the oscillating
coherent field quickly losses its energy through violent particle
production due to parametric resonance~\cite{KLS}. The parametric
resonance decay occurs during the $\phi$ oscillation. In the
present model, only a self-production (e.g. $\langle \phi \rangle +
\langle \phi \rangle \rightarrow \Phi + \Phi$) is
effective.\footnote{%%
In principle, the $\Phi$ can couple to other fields $\eta$.
Although the mass of $\eta$ is large $\sim O(F_a)$ in the true vacuum
of $\Phi$, the coherent $\Phi$ field also decays into $\eta$ particles
through the parametric resonance. However, this does not change the
conclusion in this paper as will be noted in the end of this section. }            %%
(Notice that the ``radial'' oscillation induces excitations of the
phase (=axions) as well as $\phi$.)  In this case the classical
picture does not apply at all.  The important fact is that the quantum
fluctuations due
to the emitted $\phi$ and axion particles becomes quite large.~\footnote{%%
Notice that the wavelengths of the fluctuations induced by the decay
are much shorter than the horizon scale. Therefore, these fluctuations
do not make any contribution to the fluctuations with wavelengths
relevant to large scale structures of the universe. }
Contributions of the quantum fluctuations $\delta \phi$ and $\delta a
= |\phi|\delta \theta_a$ to the energy density of $\Phi$ are $\sim
g(\delta \phi)^2M_{\rm pl}^2$~\cite{KLS} and $\sim g(\delta
\theta_a)^2M_{\rm pl}^4$,~\footnote{%%
This energy density comes from the kinetic term of the axion field. }
respectively. Since most of the initial
oscillation energy $g M_{\rm pl}^4$
is transferred to
$\Phi$ particles through the parametric resonance, we expect that the
fluctuations of $\phi$ and $\theta_a$ are given by $\delta \phi \sim
M_{\rm pl}$ and $\delta \theta_a \sim 1$, respectively, which leads to 
production of domain walls.

For more quantitative analysis, we need to solve equations for modes
of the fluctuations $\delta \phi_k$ and $\delta a_k$ which correspond
to $\phi$ particle and axion with momentum $k$, respectively. However
it is more convenient to use two field $X$ and $Y$ which are defined
by $X \equiv Re(\Phi)$ and $Y\equiv Im(\Phi)$ with initial condition
$X(0)=|\phi(0)|$ and $Y(0)=0$. Then the fluctuations of $\phi$ and
$\theta_a$ are approximately given by $\delta \phi \simeq \delta X$
and $\delta \theta_a \simeq \delta Y /|X|$. The evolutions of $X$,
$\delta X$ and $\delta Y$ are described by
\begin{eqnarray}
    \ddot{X} + g X (X^2 - F_a^2) 
       & = & - 3 g \langle (\delta X)^2 \rangle X
             -   g \langle (\delta Y)^2 \rangle X,   
    \label{eq-X}\\
    \delta\ddot{X}_k + [k^2 - g F_a^2 + 3 g X^2] \delta X_k    
       & = & - 3 g \langle (\delta X)^2 \rangle \delta X_k
             -   g \langle (\delta Y)^2 \rangle \delta X_k,
    \label{delta-X}\\
    \delta\ddot{Y}_k + [k^2 - g F_a^2 + g X^2] \delta Y_k  
       & = & -   g \langle (\delta X)^2 \rangle \delta Y_k
             - 3 g \langle (\delta Y)^2 \rangle \delta Y_k,
    \label {delta-Y}
\end{eqnarray}
where we have used the mean field approximation ($(\delta X)^3 \simeq 
3\delta X \langle (\delta X)^2 \rangle$,$\cdots$) 
and neglected the cosmological expansion.\footnote{%%
The cosmological expansion reduces the effect of
parametric resonance  in general. However we find that the reduction
effect is not large enough to avoid the conclusion in this paper.}
  Before we present results of
the full numerical integration of the above equations, we briefly
discuss instability of the equations for $\delta X$ and $\delta
Y$. When the amplitude of the $X$ oscillation is much larger than
$F_a$, a solution to eq.(\ref{eq-X}) is approximately given by 
\begin{equation}
    X \simeq Z \sin( c \sqrt{g} Z t),
\end{equation}
where $c={\cal O}(1)$ and $Z$ is the amplitude of the
oscillation. Using this 
solution and neglecting  back reactions, eqs.(\ref{delta-X}) and
(\ref{delta-Y}) become well-known Mathieu equation:
\begin{eqnarray}
    \delta X_k'' + [A_X(k) - 2q_X \cos (2z)]\delta X_k & \simeq & 0, \\
    \delta Y_k'' + [A_Y(k) - 2q_Y \cos (2z)]\delta Y_k & \simeq & 0,
\end{eqnarray}
where $A_X = \frac{k^2-g F_a^2}{c^2 g Z^2}+2q_X$, $q_X =
\frac{3}{4c^2}$, $A_Y=\frac{k^2-g F_a^2}{c^2 g Z^2}+2q_Y$,
$q_Y=\frac{1}{4c^2}$, $z = c\sqrt{g}Zt$ and a prime denotes a
derivative with respect to $z$. For $q_{X,Y} \lesssim 1$, the Mathieu
equation has instability for $k$ which satisfies $A(k)\sim 1, 4, 9,
\cdots$. Since $A_X(k) > 1$ the strongest instability (resonance) for
$\delta X_k$ occurs in the second instability band, i.e. $A_X(k)\sim
4$. On the other hand the resonance for $\delta Y_k$ occurs in the
first instability band and the instability is stronger than that for
$\delta X_k$. Since $\delta Y_k$ roughly corresponds to the axion with
momentum $k$ it is expected that the homogeneous $X$ field decays
dominantly into axions. However, as the fluctuations increase, back
reactions, i.e., terms in RHS of eqs.(\ref{delta-X}) and
(\ref{delta-Y}), become significant and the instability is weakened.
To see the evolution of the fluctuations with back reactions, we need
numerical calculations as shown below.

\begin{figure}[htbp]
\epsfxsize=15cm
\begin{center}
\leavevmode\epsffile{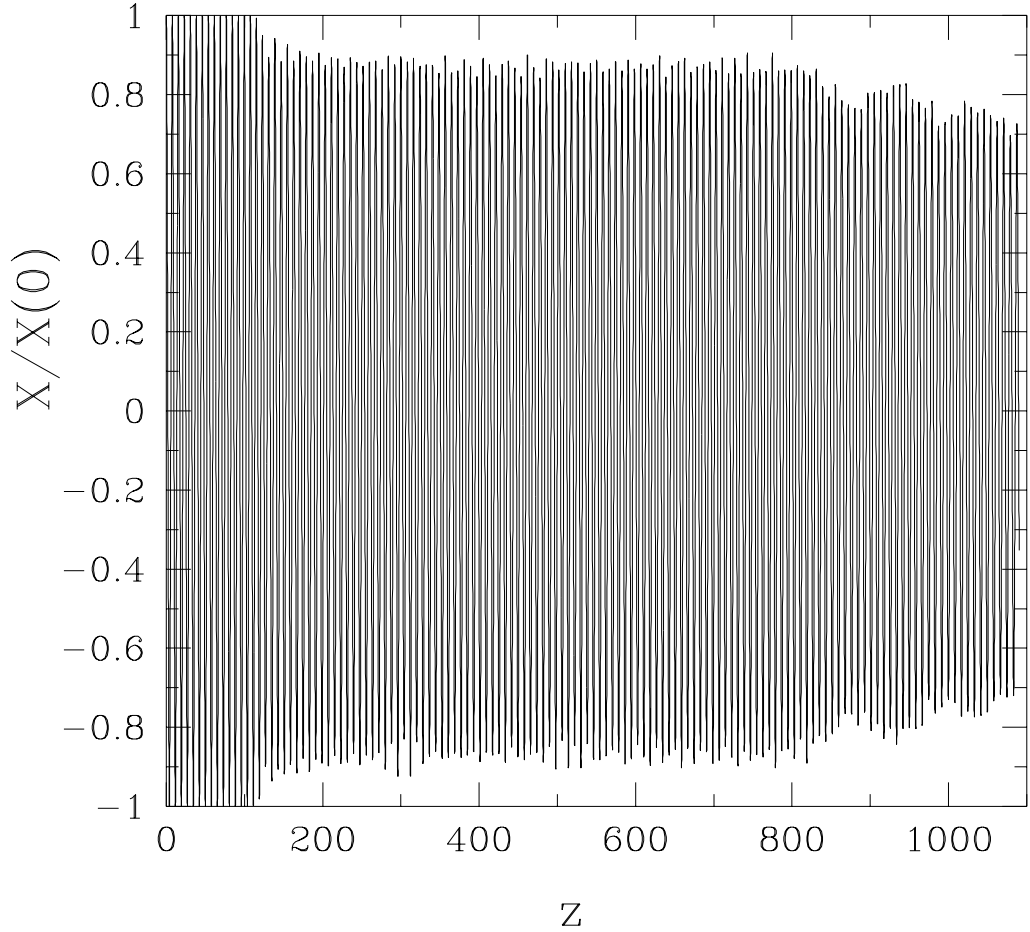}
    \caption{Time evolution of the fluctuations of $X$. $z$ is the
    time in units of $(c\protect\sqrt{g}Z(0))^{-1}$ where $c$ is the
    $O(1)$ constant, $g$ is the self-coupling constant and $Z(0)$ is
    the initial amplitude of the oscillation of the Peccei-Quinn
    scalar field. We take $X(0)=M_{\rm pl}/3$, and $g=10^{-13}$.}
    \label{fig:evol-x}
\end{center}
\end{figure}

\begin{figure}[htbp]
\epsfxsize=15cm
\begin{center}
\leavevmode\epsffile{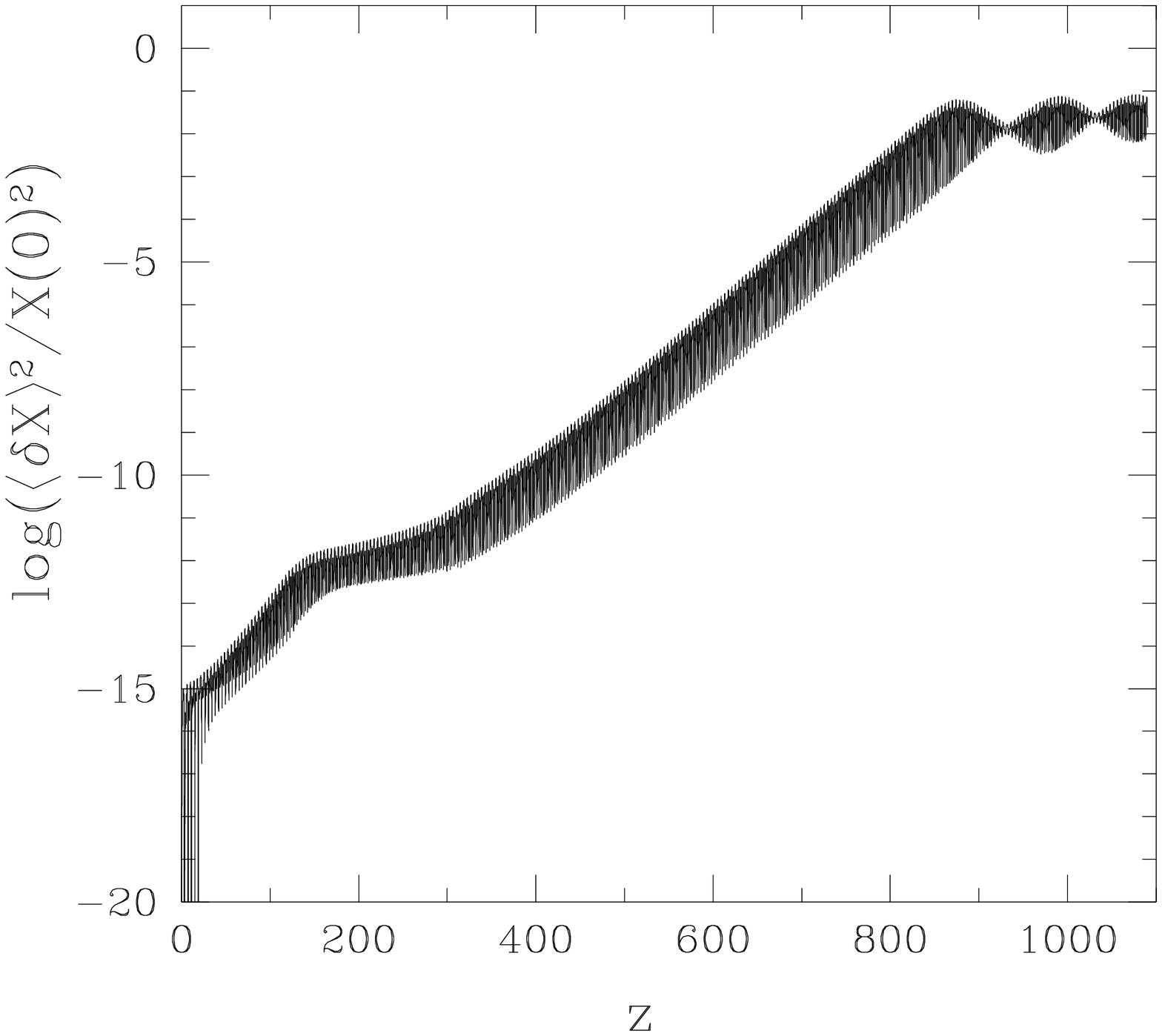}
    \caption{Time evolution of the fluctuations of $\delta X$.}
    \label{fig:evol-dx}
\end{center}
\end{figure}

\begin{figure}[htbp]
\epsfxsize=15cm
\begin{center}
\leavevmode\epsffile{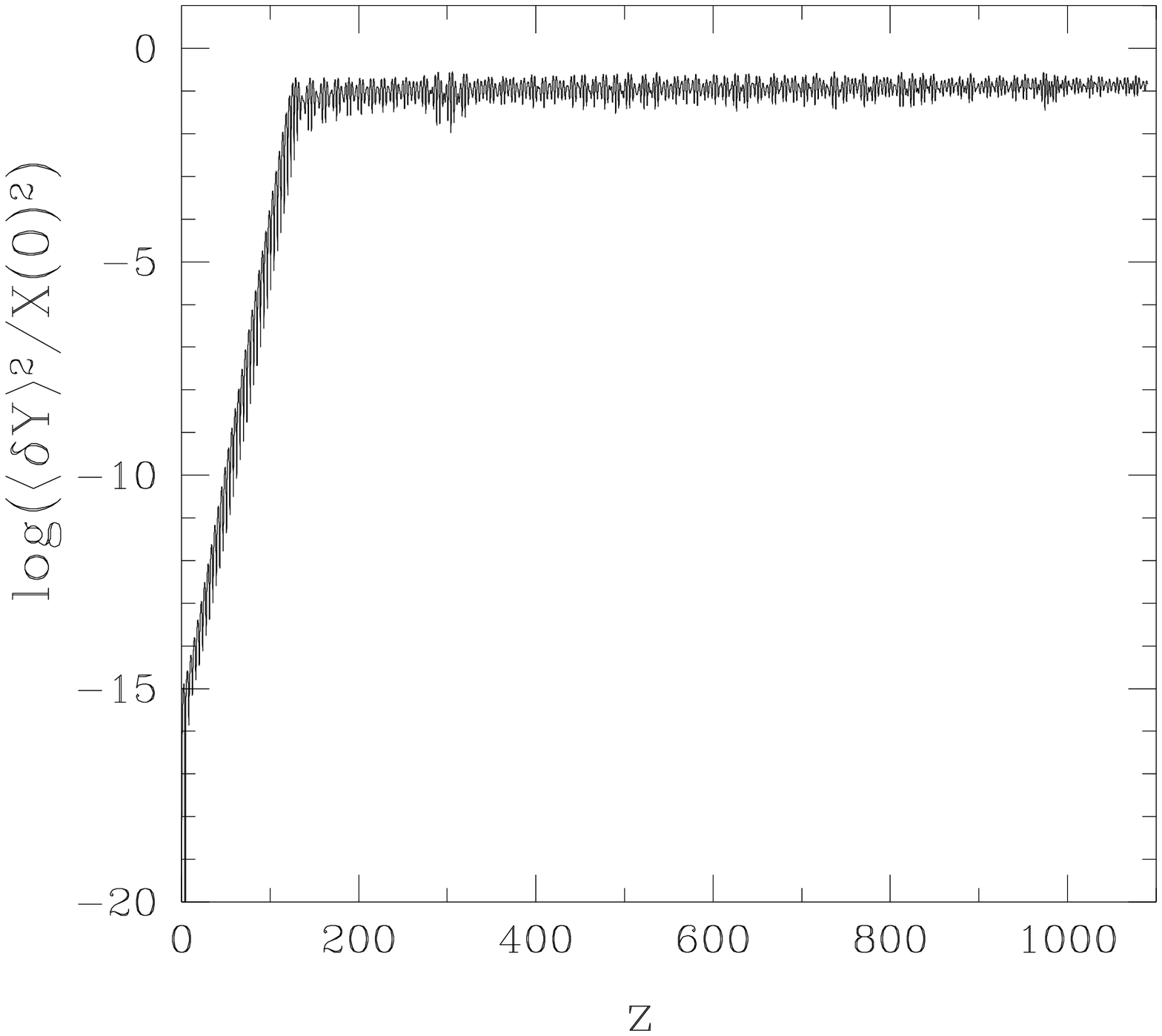}
    \caption{Time evolution of the fluctuations of $\delta Y$.}
    \label{fig:evol-dy}
\end{center}
\end{figure}

\begin{figure}[htbp]
\epsfxsize=15cm
\begin{center}
\leavevmode\epsffile{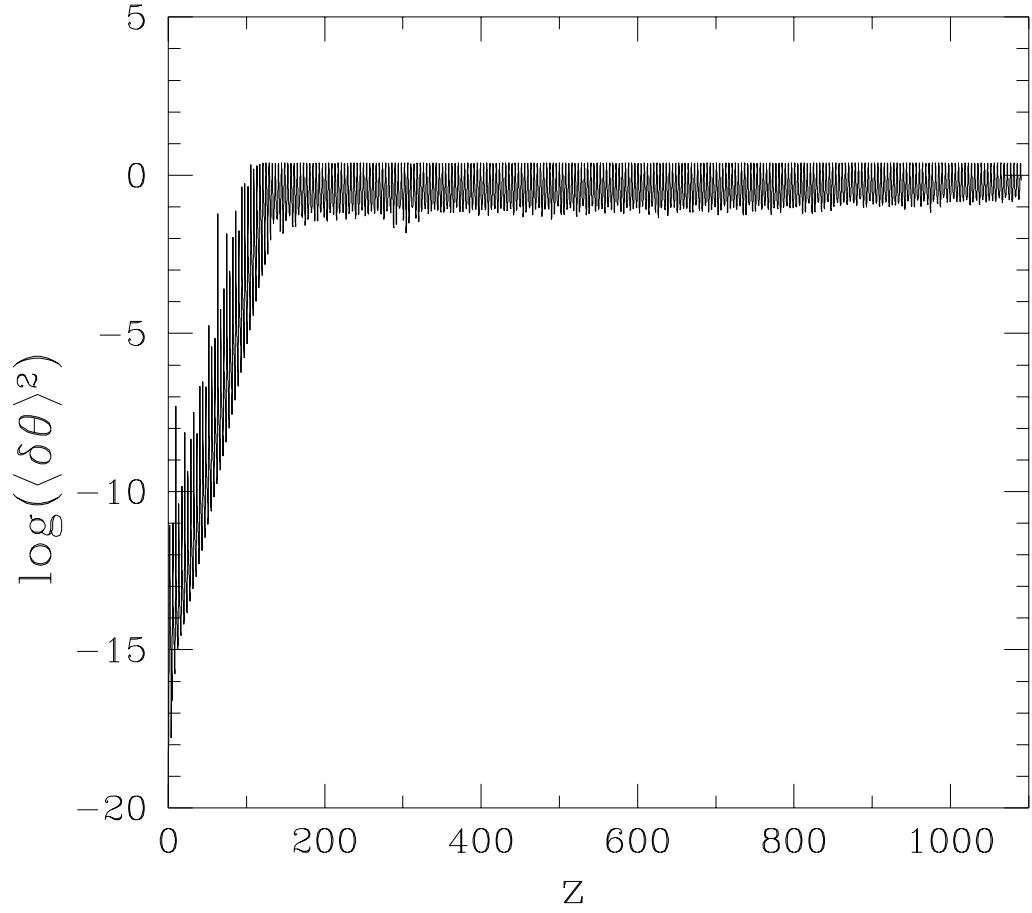}
    \caption{Time evolution of the fluctuations of $\theta_a$.We 
    define $\delta \theta$ as $\arctan(\protect\sqrt{(\delta Y)^2}/|X|)$. }
    \label{fig:evol-dth}
\end{center}
\end{figure}

Precise time evolution of the fluctuations is obtained by full
numerical integration of eqs.(\ref{eq-X}) -- (\ref{delta-Y}) and 
results are shown in Fig.~\ref{fig:evol-x} -- \ref{fig:evol-dth}. As
seen in Fig.~\ref{fig:evol-dx} and \ref{fig:evol-dy}, the fluctuations
of both $X$ and $Y$ increase through the parametric resonances. Since
the instability of $\delta Y$ is stronger, the fluctuations of $Y$
increase faster than those of $X$. As the fluctuations increase, the
back reactions become important and weaken the instability and finally
the fluctuations become almost constant. As is expected, the final
fluctuations are large and, in particular, the phase fluctuations are
of order 1 ($\delta \theta_a \sim 1$), which results in the production
of too many domain walls (see Fig.~\ref{fig:evol-dth}).

The above mechanism for producing domain walls seems quite general. In
fact we have solved equations for the fluctuations with a potential $V
= g/M_{\rm pl}^2 (|\Phi|^2-F_a^2)^3$ and obtain almost the same
result. We also consider interaction between $\Phi$ and other scalar
fields $\eta$:
${\cal L}_{int}=-f |\Phi|^2 |\eta|^2$.\footnote{%%
Possible trilinear coupling contributing to $\Phi$ decay is 
${\cal L}_{int} = -f'\Phi \eta^2$. This interaction, however, gives
unbounded potential at the vacuum $|\Phi| = F_a$ and hence we discard
this possibility.}
If the coupling $f$ is larger than $g$, the $\Phi$ field first decays
into $\eta$-particles. However, the back reactions of $\eta$-particles
suppress the resonance decay when $\delta \eta \sim
M_{pl}$~\cite{KLS}. Then the remaining $\Phi$ field may decay into
$\phi$-particles and axions through the parametric resonance, which
results in large phase fluctuations ($\sim O(1)$). Thus  the
production of $\eta$ particles cannot suppress the violent fluctuations of
the axion phase.  Therefore, the production of domain walls may not be
avoided in the chaotic inflationary universe. Since the produced
domain walls dominate the density of the universe for the axion model
with $N \ge 2$, only the model with $N=1$ is allowed. Fortunately, the
$N=1$ model is easily constructed in a hadronic axion model for
example~\cite{Kim2}.

\section{Supersymmetric Potential}
\label{sec:SUSY}

We now consider a SUSY extension of a hadronic axion model. We take
the following simple superpotential:
\begin{equation}
    \label{super-pot}
    W = h (\Psi_{+}\Psi_{-} - F_a^2)\Psi_{0},
\end{equation}
where $\Psi_{+}, \Psi_{-}$ and $\Psi_{0}$ are chiral superfields with
Peccei-Quinn charges $+1, -1$ and $0$, respectively, and the coupling
constant $h$ is
assumed to be $\sim O(1)$.\footnote{%%
We also assume a pair of quark $Q$ and antiquark $\bar{Q}$ whose
Peccei-Quinn charges are $-1/2$. This charges are chosen so that $Q$
and $\bar{Q}$ have a Yukawa coupling $W = f Q\bar{Q}\Psi_{+}$. This model
has the QCD anomaly factor $N=1$. }
Then the scalar potential is written as
\begin{equation}
    \label{susy-pot}
    V_{\rm SUSY} = h^2 |\Phi_{+} \Phi_{-} - F_a^2|^2 
    + h^2(|\Phi_{+}|^2 + |\Phi_{-}|^2)|\Phi_{0}|^2
\end{equation}
where $\Phi_{+}, \Phi_{-}$ and $\Phi_{0}$ are scalar components of
superfields $\Psi_{+}, \Psi_{-}$ and $\Psi_{0}$, respectively.  Here
we should note that there are flat directions
that satisfy
\begin{equation}
    \label{constraint}
    \Phi_{+} \Phi_{-} = F_a^2, ~~~~~ \Phi_{0}=0.
\end{equation}
We also add  soft SUSY breaking terms to the potential. Then,  the
scalar potential $V$ that we should study is
\begin{equation}
    \label{scalar-pot}
    V = V_{\rm SUSY} + m_{+}^2 |\Phi_{+}|^2 + m_{-}^2 |\Phi_{-}|^2
    +  m_{0}^2 |\Phi_{0}|^2,
\end{equation}
where $m_{\pm, 0}$ are soft masses of $O(100 {\rm GeV})$.

\begin{figure}[htbp]
\epsfxsize=15cm
\begin{center}
\leavevmode\epsffile{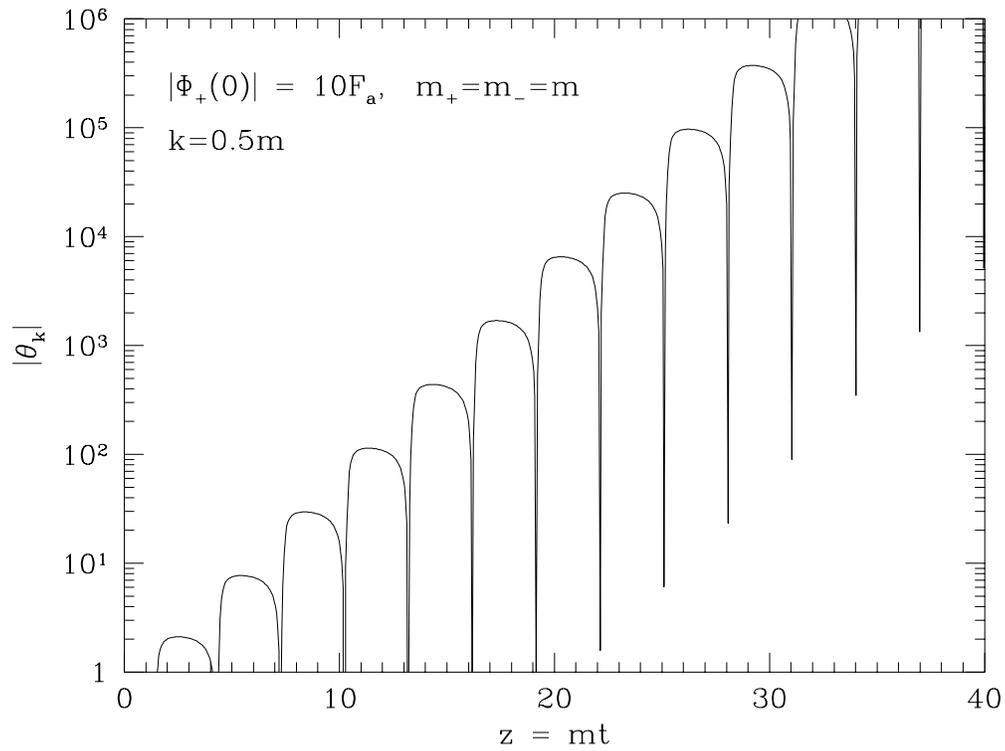}
    \caption{Time evolution of the phase fluctuations with 
  $k=0.5m_{+}$. We take $|\Phi_{+}(0)| = 10F_a$ and
  $m=m_{+}=m_{-}$. This figure shows clearly the instability of the
  phase fluctuations.}
    \label{fig:para}
\end{center}
\end{figure}

We discuss cosmological evolution of these scalar fields.  The
chaotic situation at the Planck time may lead to large initial values
of order $M_{\rm pl}$ for $\Phi_{+}, \Phi_{-}$ and $\Phi_0$. The soft
masses can be negligible at the inflation epoch since they are much
smaller than the expansion rate of the universe ($H \sim
10^{14}$~GeV).~\footnote{%%
At the inflation epoch, non-vanishing vacuum energy density breaks
SUSY and may generally give scalar masses of the order of the Hubble
constant $H$~\cite{Dine}. However, there exist a class of supergravity
models where such scalar masses are negligible compared with
$H$~\cite{Gaillard}. We assume such a class of models in this paper.}
Therefore these fields roll down into the flat direction given by
eq.(\ref{constraint}).  However, one scalar field, either $\Phi_{+}$
or $\Phi_{-}$, may stay at the initial position, i.e., $|\Phi_{+}|$ or
$|\Phi_{-}| \sim M_{\rm pl}$.  We assume that it is the case and take,
for definiteness, $|\Phi_{+}| \sim M_{\rm pl}$ and eliminate
$\Phi_{-}$ and $\Phi_0$ using eq.(\ref{constraint}).  Then, the
effective lagrangian for $\Phi_{+}$ is written as
\begin{eqnarray}
     L  & = &  L_{\rm kin} - V, \\
     \label{kinetic}
     L_{\rm kin} & = & \frac{|\Phi_{+}|^4 + F_a^4}{|\Phi_{+}|^4}
     \partial_{\mu}\Phi_{+}^{*}\partial^{\mu}\Phi_{+},\\
    \label{plus-pot}
     V  & =& m_{+}^2  |\Phi_{+}|^2 
     + m_{-}^2 F_a^4 |\Phi_{+}|^{-2}.
\end{eqnarray}
Since the potential (\ref{plus-pot}) is very steep at $\Phi_{+} =0$
and the kinetic term is always positive, the energy conservation never
allows the $\Phi_{+}$ to pass through the origin $(\Phi_{+}=0)$. Thus,
this potential has effectively only one minimum. From the above
lagrangian we obtain equations of motion for the homogeneous field
$\Phi_{+}$ and phase fluctuations $\delta \theta$:
\begin{eqnarray}
    \varphi'' - 2\frac{\varphi'^{2}}{(\varphi^4 + 1)\varphi}
    + \frac{\varphi^4 - (m_{+}/m_{-})^2}{\varphi^4 + 1}\varphi & = & 0,
    \label{radial-evolv}\\
    \theta_k'' + k^2\theta_k 
    + \frac{2(\varphi-1)}{\varphi(\varphi+1)}\varphi' \theta_k' & = &
    0,\label{mode-evolv}
\end{eqnarray}
where the prime denotes $d/(m_{+}dt)$, $\varphi \equiv |\Phi_{+}|/F_a$
and $\theta_k \sim \int dx^3 \delta\theta e^{ikx}$. In
eq.(\ref{radial-evolv}) the back reactions of $\delta\theta$ are
assumed to be neglected. Although eq.(\ref{mode-evolv}) looks
different from the Mathieu equation,  numerical calculations show that 
it has indeed strong instability (see Fig.~\ref{fig:para}). Therefore
we expect that the parametric resonance induces large phase
fluctuations and results in the production of domain walls. However
these domain walls are cosmologically harmless since we consider the
hadronic axion model with $N=1$ as discussed in the previous section.

Let us now estimate isocurvature fluctuations of the axion in the SUSY
model.  First we rewrite $\Phi_{+}$ and $\Phi_{-}$ as
\begin{equation}
    \Phi_{\pm} = v_{\pm}
    \exp{i\frac{a_{\pm}}{\sqrt{2}|v_{\pm}|}}, ~~~~~~
    -\infty < v_{\pm} < \infty. 
\end{equation}
We also define the fields $a$ and $b$ by
\begin{eqnarray}
    a & = & \frac{v_{+}}{(v_{+}^2 + v_{-}^2)^{1/2}}a_{+}
     -  \frac{v_{-}}{(v_{+}^2 + v_{-}^2)^{2}}a_{-}, \\
    b & = & \frac{v_{-}}{(v_{+}^2 + v_{-}^2)^{1/2}}a_{+}
     +  \frac{v_{+}}{(v_{+}^2 + v_{-}^2)^{1/2}}a_{-}.
\end{eqnarray}
 From eq.(\ref{susy-pot}), the potential $V_{b}$ for $b$ is given by
\begin{equation}
    V_b = -2 h^2 F_a^2 v_{+}v_{-} \cos 
    \left( \frac{(v_{+}^2 + v_{-}^2)^{1/2}}{\sqrt{2}v_{+}v_{-}}b\right).
\end{equation}
Thus the mass of $b$ is $\sim (v_{+}^2 + v_{-}^2)^{1/2} \sim
M_{\rm pl} \sim 10^{19} {\rm GeV}$, which is much larger than the
Hubble constant during the inflation and hence $b$ quickly becomes zero
at the inflation epoch. On the other hand, the potential for $a$ is flat
and is regarded as the axion. At the inflation epoch, the quantum fluctuations of $a$ are given by, assuming $v_{+} \gg
v_{-}$,
\begin{equation}
    \delta a \simeq  \delta a_{+} 
    \simeq \frac{H}{2\pi}.
\end{equation}
Since the mass of $b$ is very large compared with $H$, the quantum
fluctuations of $b$ are negligible and hence $\delta a_{-} \simeq
-(v_{-}/v_{+})\delta a_{+}$.  Therefore, the fluctuations for phases
($\theta_{\pm}$) of $\Phi_{\pm}$ are given by
\begin{equation}
    \delta \theta_{\pm} \equiv \frac{\delta a_{\pm}}{\sqrt{2}v_{\pm}} 
    \simeq  \pm \frac{H}{2\sqrt{2}\pi v_{+}}.
    \label{phase-flu}
\end{equation}
Since $v_{+} \sim M_{\rm pl}$ and $H\simeq 10^{14}$~GeV, fluctuations
of the phase $\theta_{\pm}$ are quite small ($\delta \theta_{\pm}\sim
10^{-6} - 10^{-5}$).  After the axion obtains a mass, the phase
fluctuations result in isocurvature density fluctuations and produce
the CBR anisotropies as explained in the non-SUSY case.  When
$\Phi_{+}$ and $\Phi_{-}$ settle down to the true minimum of the
potential (i.e., $v_{+} \sim v_{-} \sim F_a$), the axion field is
written as $a \sim a_{+} - a_{-}$ and its fluctuation $\delta a$ is
given by $\delta a \sim F_a \delta \theta_{+}$, which gives $\delta
\rho_a/\rho_a \sim \delta \theta_{+}/\theta_{+}$. Thus the $\delta
T/T$ is predicted as $10^{-6} - 10^{-5}$, which is consistent with the
COBE observation.  In particular, if $F_a \sim 10^{12}$~GeV, the
predicted CBR anisotropies due to the isocurvature density
fluctuations are large enough to be detected by future observations as
stressed in ref.~\cite{KSY}. 

In SUSY Peccei-Quinn models, there exists an axino which is a
supersymmetric partner of the axion. The mass of the axino is presumed
to be the SUSY scale, i.e. $100$~GeV -- $1$~TeV. The axino of such a
large mass is cosmologically harmless since it decays into lighter
particles quickly.

\section{Conclusion}

In summary, we have investigated two cosmological axion problems
(isocurvature fluctuations and domain-wall formation) in chaotic
inflationary universe. The isocurvature-fluctuation problem is solved
if potential for the Peccei-Quinn scalar field is very flat.  However,
we have found that too many domain walls are produced through the
parametric resonance decay of the Peccei-Quinn scalar field and only
the model with $N=1$ is allowed.  We have also pointed out that the
flat potential which is necessary to solve the
isocurvature-fluctuation problem is naturally obtained in a SUSY
extension of the Peccei-Quinn model. Namely, we do not need any small
coupling constant in the SUSY model for obtaining such a flat
potential. On the other hand a very small coupling constant ($g
\lesssim 10^{-11}$) is necessary in non SUSY Peccei-Quinn models.

\newpage

\begin{center}
    \textbf{Figure Captions}
\end{center}

\noindent
\begin{description}
%%%
\item[Fig.1] Time evolution of the fluctuations of $X$. $z$ is the
    time in units of $(c\protect\sqrt{g}Z(0))^{-1}$ where $c$ is the
    $O(1)$ constant, $g$ is the self-coupling constant and $Z(0)$ is
    the initial amplitude of the oscillation of the Peccei-Quinn
    scalar field. We take $X(0)=M_{\rm pl}/3$, and $g=10^{-13}$.
%%%
\item[Fig.2] Time evolution of the fluctuations of $\delta X$.
%%%
\item[Fig.3] Time evolution of the fluctuations of $\delta Y$.
%%%
\item[Fig.4] Time evolution of the fluctuations of $\theta_a$.We 
    define $\delta \theta$ as $\arctan(\protect\sqrt{(\delta
    Y)^2}/|X|)$.
%%%
  \item[Fig.5] Time evolution of the phase fluctuations with
    $k=0.5m_{+}$. We take $|\Phi_{+}(0)| = 10F_a$ and $m=m_{+}=m_{-}$.
    This figure shows clearly the instability of the phase
    fluctuations.
\end{description}

\newpage

\epsfverbosetrue

\newpage

\newpage

\newpage

\newpage

\epsfverbosefalse

\end{document}